\documentclass[twocolumn,showpacs,fleqn,nobibnotes]{revtex4}

\usepackage{amsmath}
\usepackage{graphicx}
\usepackage{float}
\usepackage{subfigure}

\def\lsim{\raise0.3ex\hbox{$<$\kern-0.75em\raise-1.1ex\hbox{$\sim$}}}
\def\gsim{\raise0.3ex\hbox{$>$\kern-0.75em\raise-1.1ex\hbox{$\sim$}}}

\renewcommand{\vec}[1]{\boldsymbol{#1}}
\newcommand{\dif}{\mathrm{d}}

\begin{document}

\title{Diffractive photoproduction of $Z^0$ bosons in coherent interactions
at  CERN-LHC}
\pacs{12.38.-t;12.38.Bx;14.70.Hp}
\author{V.P. Gon\c{c}alves$^{a}$ and M.V.T. Machado$^b$}
\affiliation{$^a$ Instituto de F\'{\i}sica e Matem\'atica, Universidade Federal de Pelotas\\
Caixa Postal 354, CEP 96010-090, Pelotas, RS, Brazil\\
$^b$ Centro de Ci\^encias Exatas e Tecnol\'ogicas, Universidade Federal do Pampa \\
Campus de Bag\'e, Rua Carlos Barbosa. CEP 96400-970. Bag\'e, RS, Brazil}

\begin{abstract}

The exclusive $Z^0$ photoproduction  at high energies in $\gamma p (A)$, $p p$ and $AA$ collisions is investigate within the color dipole formalism. We generalize the description of the deeply virtual compton scattering (DVCS) process, which describe quite well the HERA data,  for the production of $Z^0$ bosons and estimate the  total cross section for the exclusive process $\gamma^* h\rightarrow Z^0 h$ ($h=p,A$) for different  energies, photon virtualities and atomic numbers. As hadrons at collider energies are a source of
Weizs\"{a}cker - Williams photons, we consider electromagnetic interactions in hadron-hadron collisions at Tevatron and LHC energies and estimate the rapidity distribution and total cross section for $Z^0$ production in the  $h\,h\rightarrow h\, Z^0 \,h$ process. This is the first estimation for such a  process in literature. It can allow us to study, for instance, the physics of hadronic $Z^0$ decays in a clean environment characterized by two rapidity gaps. Our results indicate that the experimental analyzes of this process could be feasible in $pp$ but the physics scenario for $AA$ collisions is not promising.

\end{abstract}

\maketitle

\section{Introduction}

 The Large Hadron Collider (LHC) at CERN will explore physics at TeV scale, opening a new territory where ground-breaking discoveries are expected \cite{ellis}. One primary goal of the LHC is the search and study of the properties of the Higgs boson, which is  motivated by  the determination of the electroweak symmetry breaking  mechanism and discrimination among the models for the physics beyond of the Standard Model (SM). Recently, there has been a great deal of attention devoted to study the central exclusive diffractive (CED) Higgs boson production, $h_1 h_2 \rightarrow h_1 \otimes H \otimes h_2$, where $\otimes$ is a rapidity gap in the final state, characterized by no activity between the outgoing hadrons and the decay products of the Higgs boson, i.e. by a clean experimental signature (For a recent review see, e.g., Ref. \cite{forshaw}).

Another  interesting physics signal as well as an important background to a number of processes indicating the presence of new physics is the production of the $Z^0$ boson \cite{reviewZ0}. It should be produced copiously in $pp$ and $AA$ collisions at LHC, decaying after into lepton pairs or hadrons  (See, e.g., Ref. \cite{martin}). However, the hadronic $Z^0$ decays are expected to be difficult to identify, due to the overwhelming   background of QCD multi-jet production, owing to the hadronic event environment.  It implies that in $pp$ collisions it is almost to impossible to study the physics of hadronic $Z^0$ decays. Consequently, although  the LHC can be considered a $Z^0$ factory, it only allows for high-statistics measurement in final states with leptons.  In this letter we propose an alternative to produce $Z^0$ bosons and study the related  physics from hadronic $Z^0$ decays in a clean environment characterized by two rapidity gaps between the $Z^0$ and the outgoing hadrons which can be detected in the forward proton detectors proposed to be installed in the CMS and ATLAS collaborations at LHC.  Basically, we propose to study the diffractive $Z^0$ photoproduction in coherent hadron - hadron interactions. This process is characterized by the photon - hadron interaction, with the photon stemming from the electromagnetic field
of one of the two colliding hadrons (For recent reviews see Ref. \cite{upcs}). The $Z^0$ is produced when the photon fluctuates into a quark - antiquark ($q \bar{q}$) pair, scatters diffractively on the  target hadron through gluon exchanges and emerges as a $Z^0$. The $Z^0$ mass provides the hard scale which justifies a perturbative calculation. The total cross section for the $h\,h\rightarrow h\otimes Z^0 \otimes h$ process is given by
\begin{eqnarray}
\sigma (h_1 h_2 \rightarrow h_1 Z^0 h_2)\, = \int \limits_{\omega_{min}}^{\infty} d\omega \frac{dN_{\gamma}(\omega)}{d\omega}\,\sigma_{\gamma h \rightarrow Z^0 h} \left(W_{\gamma h}^2\right)\,,
\label{sigAA}
\end{eqnarray}
where
$\omega$ is the photon energy   and $ \frac{dN_{\gamma}(\omega)}{d\omega}$ is the equivalent flux of photons from a charged hadron. Moreover, $\omega_{min}=M_{Z^0}^2/4\gamma_L m_p$, $\gamma_L$ is the Lorentz boost  of a single beam,  $W_{\gamma h}^2=2\,\omega\sqrt{S_{\mathrm{NN}}}$  and
$\sqrt{S_{\mathrm{NN}}}$ is  the c.m.s energy of the
hadron-hadron system \cite{upcs}. This process is characterized by small momentum transfer and energy loss, which implies that the outgoing hadrons should be detected in the forward regions of the main LHC detectors.
The study of this process is feasible considering the proton tagging detectors (Roman Pots) already planned for the initial start-up of the LHC in the 220/240 m region by CMS/TOTEM and ATLAS. Moreover, the situation can still be improved in the future if the Roman Pots at 420 m from the interaction points of ATLAS and/or CMS were installed (For details see, e.g., Ref. \cite{detec}). 

One have that the cross section for $Z^0$ production, Eq. (\ref{sigAA}), is directly dependent of the magnitude and energy dependence of the $\gamma h \rightarrow Z^0 h$ cross section, which is currently unknown. Here, we estimate this cross section considering a generalization of the dipole picture approach \cite{dipole} currently used for the description of the  deeply virtual Compton scattering (DVCS) in $ep$ collisions at  HERA \cite{dvcs,fss,FM,MPS}.  Specifically, in this approach the exclusive process $\gamma\, p\rightarrow E\,p$ is described  by the interaction of $q\bar{q}$ pairs (color dipoles), in which the virtual boson fluctuate into, with the nucleon. The scattering amplitude is given by the convolution of the dipole-nucleon cross section with the overlap of the virtual incoming boson and outgoing exclusive ($E$) final state wave functions. This approach provides a very good description of the data on $\gamma p$ inclusive production, $\gamma \gamma$ processes, diffractive deep inelastic and vector meson production (See, e.g., Ref. \cite{fss}). As the wave functions and dipole-target cross section are reasonably well known, one have that our calculation of the diffractive $Z^0$ photoproduction introduces no new parameters.  Therefore, the $Z^0$ production in $\gamma h$ interactions can be considered as an important test of the dipole approach formalism as well as a search of information about the dipole - target cross section and, consequently, of the QCD dynamics at high energies.


\section{Diffractive production of $Z^0$ at high energies}

Let us introduce the main formulas concerning the color dipole picture applied to the exclusive processes in deep inelastic scattering (DIS). In the dipole model \cite{dipole}, the DIS is viewed as the interaction of a color dipole, i.e. mostly a quark--antiquark pair, with the target.  The transverse size of the pair is denoted by $\vec{r}$ and a quark carries a fraction $z$ of the photon's light-cone momentum.  In the target rest frame, the dipole lifetime is much longer than the lifetime of its interaction with the target.  Therefore, the elastic $\gamma^* h$ scattering is assumed to proceed in three stages: first the incoming virtual photon fluctuates into a quark--antiquark pair, then the $q\bar{q}$ pair scatters elastically on the target, and finally the $q\bar{q}$ pair recombines to form the exclusive final state (real photons, vector mesons, flavor mesons, etc). The amplitude for the elastic process $\gamma^* h\rightarrow \gamma^* h$, $\mathcal{A}^{\gamma^*h}(x,Q,\Delta)$, is simply the product of amplitudes of these three subprocesses integrated over the dipole variables $\vec{r}$ and $z$:
\begin{eqnarray}
  \mathcal{A}^{\gamma^*h}(x,Q,\Delta) & = & \sum_f \sum_{n,\bar n} \int\!\dif^2\vec{r}\,\int_0^1\!\dif{z}\,\Psi^*_{n\bar n}(r,z,Q)\nonumber \\
& \times & \mathcal{A}_{q\bar q}(x,r,\Delta)\,\Psi_{n\bar n}(r,z,Q),
  \label{eq:elamp}
\end{eqnarray}
where $\Psi_{n\bar n}(r,z,Q)$ denotes the amplitude for the incoming virtual photon to fluctuate into a quark--antiquark dipole with helicities $n$ and $\bar n$ and flavor $f$. The quantity $\mathcal{A}_{q\bar q}(x,r,\Delta)$ is the elementary amplitude for the scattering of a dipole of size $\vec{r}$ on the proton, $\vec{\Delta}$ denotes the transverse momentum lost by the outgoing proton (with $t=-\Delta^2$), $x$ is the Bjorken variable and $Q^2$ is the photon virtuality.  Eq. (\ref{eq:elamp}) has been extensively used to compute the $\gamma^*p$ cross section and proton structure functions as well (See, e.g.,  Refs. \cite{fss,kowtea}). On the other hand, the amplitude for production an exclusive final state, $E$, is  given by \cite{kowtea}
\begin{eqnarray}
 \mathcal{A}^{\gamma^* h\rightarrow E h}(x,Q,\Delta) = \int\!\dif^2\vec{r}\int_0^1\!\dif{z}\,(\Psi_{E}^{*}\Psi_{\gamma^*})\,\mathcal{A}_{q\bar q}(x,r,\Delta),
  \label{eq:ampvecm}
\end{eqnarray}
where $(\Psi^*_{E}\Psi_{\gamma^*})$ denotes the overlap of the virtual incoming photon and outgoing exclusive final state $E$ wave functions (non-forward wave functions will be used \cite{bartels_non}).  Consequently, the elastic diffractive cross section is then given by
\begin{eqnarray}
  \frac{\dif\sigma^{\gamma^* h\rightarrow E h}}{\dif t}
  & = & \frac{1}{16\pi}\left\lvert\mathcal{A}^{\gamma^* h\rightarrow E h}(x,Q,\Delta)\right\rvert^2
  \label{eq:xvecm1}
\end{eqnarray}
For the elementary dipole-proton amplitude, $\mathcal{A}_{q\bar q}(x,r,\Delta)$, we will consider the saturation model proposed in Ref. \cite{MPS}, which describe quite well the current HERA data for exclusive processes.  It is based on saturation physics, which predicts the
limitation on the maximum phase-space parton density that can be
reached in the hadron wave function (parton saturation) and  the
transition between the linear and nonlinear regimes of the QCD dynamics at high energies. This transition is specified  by a typical scale, which is energy
dependent and is called saturation scale $Q_{\mathrm{sat}}$  (For a review see Ref. \cite{hdqcd}).
In the model considered the elementary dipole-proton amplitude is given by \cite{MPS}
\begin{eqnarray}
\label{sigdipt}
\mathcal{A}_{q\bar q}(x,r,\Delta)= 2\pi R_p^2\,F(\Delta) \, N\left(rQ_{\mathrm{sat},p}(x,|t|),x\right),
\end{eqnarray}
which is an extension of the forward model \cite{Iancu:2003ge} including the QCD
predictions for non-zero momentum transfer, which reproduces the
initial model for $|t|=0$ and ensures that the saturation scale has the
correct asymptotic behaviors. The dipole-nucleon scattering amplitude at $t=0$ is given by:
\begin{equation}
\label{eq:bcgc}
 N(x,\,r) =\begin{cases}
  \mathcal{N}_0\left(\frac{rQ_{\mathrm{sat}}}{2}\right)^{2\left(\gamma_s+\frac{1}{\kappa\lambda Y}\ln\frac{2}{rQ_{\mathrm{sat}}}\right)} & :\quad rQ_{\mathrm{sat}}\le 2\\
  1-\mathrm{e}^{-A\ln^2(BrQ_{\mathrm{sat}})} & :\quad rQ_{\mathrm{sat}}>2
  \end{cases},
\end{equation}
where $Y=\ln(1/x)$, and the parameters are $\mathcal{N}_0=0.7$, $\kappa = 9.9$ and $\gamma_s=0.63$. The constants $A$ and $B$ are obtained from continuity conditions at  $rQ_{\mathrm{sat}}=2$. The $t$ dependence of the saturation scale is parameterized as $Q_{\mathrm{sat},p}^2\,(x,|t|)=Q_0^2(1+c|t|)\,x^{-\lambda}$,
in order to interpolate smoothly between the small and intermediate transfer
regions. The form factor $F(\Delta)=\exp(-B|t|)$ catches the transfer dependence of the proton vertex, which is factorized from the projectile vertices and  does not spoil the geometric scaling properties. Finally, the scaling function $N$ is obtained from the forward saturation model \cite{Iancu:2003ge}.  The remaining parameters are $Q_0=0.206$ GeV, $\lambda=0.253$,  $c=3.776$ GeV$^{-2}$ and $B=3.74$ GeV$^{-2}$.

In the case investigated here, we have the exclusive production of $Z$ at final state, $E=Z^0$. Therefore, the non-forward scattering amplitude will be given by Eq. (\ref{eq:ampvecm}), with the overlap of the incoming virtual photon and outgoing $Z^0$ final state wave functions, $(\Psi_{Z^0}^*\Psi_{\gamma^*})$. Regarding the wave functions, the $Z^0$ bosons behave in the same way as the photon, which implies that we could use the result from Ref. \cite{bartels_non}, and only the magnitude of the coupling to the quarks depends on the nature of the boson \cite{vacca}. The light-cone wavefunctions for a virtual $Z$ boson can obtained from the photon wavefunction by the replacement $e_Q\gamma_{\mu}\rightarrow (g_W/2\cos \theta_W)\gamma_{\mu}(c_V-c_A\gamma_5)$. For sake of simplicity, we consider space-like kinematics.  More recently, the exclusive $Z^0$ photoproduction has been computed taking a similar approach \cite{motyka}. They have used the correct time-like kinematics for the heavy boson photoproduction, but this correction represents numerically a small correction to the cross section. The main deviation come from the distinct saturation model considered and a different overall normalization. We realized that our original calculation has an erroneous additional factor of $2\pi$, which is corrected in the present version \cite{motyka_disc}. Fortunately, our general conclusions on the feasibility of measurement of this exclusive process remain unchanged.

Some comments are in order at this point. Firstly, it is important to emphasize that due to the large mass of the $Z^0$, the $q \overline{q}$ dipole is spatially compact, which implies that the cross section is dominated by small pair separations, i.e. by the linear regime of the QCD dynamics. That is, the mean color dipole radius is of order $r\propto 1/m_Z \ll 1/Q_{sat}$.  Concerning the saturation effects, its contribution is more  sizeable at large dipole configurations, meaning that for $Z^0$ production we should expect that the associated modifications  will be small. This is corroborated by the numerical calculations presented in what follows, where the effective energy power law for $Z$ production is of order $\lambda_{\mathrm{eff}}\simeq 0.5$. Such an effective power is clearly consistent with a BFKL-like QCD dynamics for the process and gluon resummation effects are not strongly important. Our calculation is consistent with BFKL dynamics as the employed saturation model of Ref. \cite{Iancu:2003ge} reproduces the BFKL QCD regime for small color dipoles. Since such an approach  provides a very good description of the diffractive and inclusive HERA data, including the DVCS experimental data, and the description does not include any new parameter, one believe that our predictions  are reasonable for the LHC energy.

Secondly, it is timely to discuss related approaches for the present scattering process and in what extent they could differ from the current calculation. We quote the pioneering work of Ref. \cite{BartelsLoewe} (denoted BL hereafter), where  an analysis using non-forward QCD planar ladder diagrams have been considered to compute the diffractive $Z^0$-production in an $ep$ collider. The computation allows QCD evolution in the low-$x$ region; however, a direct comparison to ours results is a hard task. As an estimation, it is found a cross section $\sigma \approx 10^{-37}$ cm$^{2}$ after including branching ratio for the $\mu$-decay \cite{BartelsLoewe}. This means a cross section $\sigma_{BL} \,(ep\rightarrow Z^0\,ep)\approx $ 3 pb, which leads that $\sigma_{BL}(\gamma p\rightarrow Z^0 p)$ is about $0.01$ pb at $\sqrt{s}=300$ GeV, consistent with the present work.  In Ref. \cite{Pumplin}, the two-gluon exchange model of the Pomeron is used to compute the $Z^0$ photoproduction, where reggeization of the gluons and interactions between them (ladder diagrams) are ignored. The main uncertainties in that calculation are the parameterization for the gluon-proton amplitude and the finite gluon mass included in the propagators to suppress contributions from long distance. The integrated cross section (energy independent) is $\sigma_{2g}(\gamma p\rightarrow Z^0 p) \simeq 0.025$ pb, with an uncertainty of a factor 2 \cite{Pumplin}. This value is consistent with our calculations for $\sqrt{s}\gg m_Z$. Finally, it would be interesting to compare this work with a conventional partonic description of timelike Compton scattering. Presently, the photoproduction of a heavy timelike photon which decays into a lepton pair, $\gamma p \rightarrow l^+l^-\,p$, is computed to leading twist and at Born level \cite{Diehl}. In our case, the virtual photon in final state is replaced by the weak neutral boson. The amplitude is given by the convolution of hard scattering coefficients, calculable in perturbation theory, and generalized parton distributions (GPDs), which describe the nonperturbative physics of the process. For instance, in Ref. \cite{Diehl}, the quark handbag diagrams (leading order in $\alpha_s$) and simple models of the relevant GPDs are used to estimate the cross section and the angular asymmetries for lepton pair production. In the coherent case (real photons from proton or nuclei) studied here, the background from Bethe-Heitler process can be completely disregarded. We remark that such a calculation can not be compared with ours since an explicit analysis using ${\cal{O}}(\alpha_s)$ accuracy, one-loop corrections to the quark handbag diagrams and other diagrams involving gluon distributions is currently unknown.

  We are now ready to compute the total cross section. In Fig. \ref{fig:1} we present the calculation for the total cross section (integrated over $|t|\leq 1$) as a function of energy for distinct photon virtualities. The range of energy is extrapolated up to $W_{\gamma p}=1$ TeV. It is verified that the cross section is weakly dependent on the virtuality, due to large mass of the $Z^0$, which determines the hard scale in combination with  $Q^2$. It is only for very large $Q^2$ that the latter scale determines the behavior of the cross section. In order to account for correct behavior near the $Z^0$ threshold production region, we have multiplied the total cross section by a factor
 $(1-\bar{x})^5$, where $\bar{x}=(M_{Z^0}+m_p)^2/W^2$. Notice that the photoproduction cross section for $W\gg M_{Z^0}$ can be parameterized as $\sigma_{tot}(\gamma p\rightarrow Z^0 p)\approx 0.3\times 10^{-4} fb \,(W/\mathrm{GeV})^{1.9}$. We emphasize that the magnitude of this cross section implies that the study of the $Z^0$ production is a hard task  in the next generation of $ep$ colliders. Let us now discuss the uncertainties on the present approach. As we have considered a generalization of the color dipole formalism for the scattering process, the main uncertainties come from the considered phenomenological (saturation) model parameters ($\lambda$, $\sigma_0$, $\gamma_s$), which we have taken from fits to $ep$ HERA data. These parameters are well determined from collider data and differs by a few percents  in a comparison among recent implementations of dipole cross section, Eq. ({\ref{eq:bcgc}). The dipole picture is particularly suitable for a qualitative analysis as the physical interpretation is clear in this representation. For instance, taking Eq. (\ref{eq:ampvecm}) and the explicit expression for the overlap of wavefunctions, we see that the main contribution of dipole transverse size comes from the interval $0\leq r \leq 2/M_Z$. This fact simplifies the integration on longitudinal momentum $z$ and the process is dominated by the so called symmetric dipole configurations. If the characteristic size of the $q\bar{q}$-pair is much smaller than the mean distance between partons, $R_0(x)=1/Q_{sat}$, the forward scattering amplitude then reads,
\begin{eqnarray}
& &  \mathcal{A}^{\gamma^* h\rightarrow Z^0 h} = \int_0^{\infty}\!\dif^2\vec{r}\int_0^1\!\dif{z}\,(\Psi_{Z^0}^{*}\Psi_{\gamma^*})\,\mathcal{A}_{q\bar q}(x,r,\Delta=0)\nonumber\\
& & \propto \int_0^{\frac{4}{M_Z^2}}\!\dif \vec{r}^2\,\frac{N(x,\vec{r}^2)}{\vec{r}^2}\approx \int_0^{\frac{4}{M_Z^2}}\!\frac{\dif \vec{r}^2}{\vec{r}^2}\left(\frac{r^2Q_{\mathrm{sat}}^2}{4}\right)^{\gamma_{\mathrm{eff}}}\,,
  \label{eq:approc}
\end{eqnarray}
where, $\gamma_{\mathrm{eff}}=\gamma_s+\frac{1}{\kappa\lambda Y}\ln\frac{2}{rQ_{\mathrm{sat}}}$. The solution for the integral above is given by,
\begin{eqnarray}
\mathcal{A}^{\gamma h\rightarrow Z^0 h}(x,\Delta=0)\propto \sqrt{2\kappa \lambda Y}\left(\frac{1}{x}\right)^{\frac{\gamma_s^2\kappa \lambda}{2}}\!\!\left[1+\mathrm{Erf}(\eta) \right],
 \label{eq:approc2}
\end{eqnarray}
with $\eta = \sqrt{1/2\kappa \lambda Y}[\ln(Q_{sat}^2/M_Z^2)-\gamma_s]$. Therefore, the effective energy power for the amplitude is given by $\lambda_{\mathrm{eff}}=\gamma_s^2\kappa \lambda /2$. Using the parameters in model in Ref. \cite{Iancu:2003ge} one has $\lambda_{\mathrm{eff}}\simeq 0.5$, which it is consistent  with the value obtained in our full numerical calculation.

\begin{figure}[t]
\includegraphics[scale=0.45]{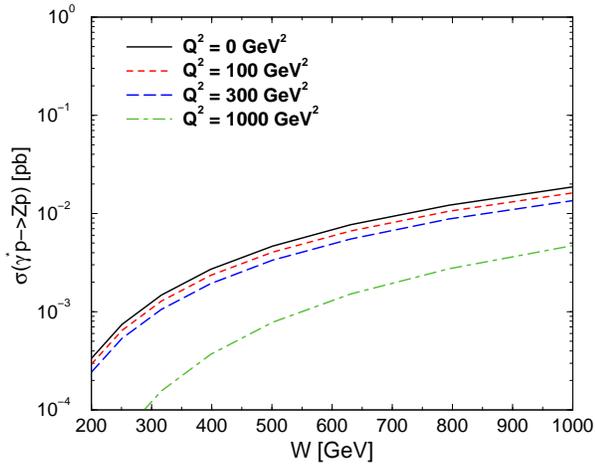}
\caption{(Color online) The total cross section (integrated over $|t|\leq 1$) for diffractive $Z^0$ production in DIS as a function of energy for several photon virtualities (see text).}
\label{fig:1}
\end{figure}

Finally, we discuss the case of the photonuclear reaction $\gamma A\rightarrow Z^0 A$. A simple assumption would be to use the forward scattering amplitude scaling with the number of nucleons, $A$, due to no absorption occuring for dipoles with very small transverse size.  For this purpose  we will rely on the geometric scaling property  of the saturation models within the color dipole approach \cite{hdqcd}.  Following the procedure from Ref. \cite {Armesto_scal}, which successfully describes small-$x$ data for $ep$ and $eA$ scattering,  we replace $R_p \rightarrow R_A$  and $Q_{\mathrm{sat},p}^2 \rightarrow (A\,R_p^2/R_A^2)^{\epsilon}\,Q_{\mathrm{sat},p}^2$ in Eq. (\ref{sigdipt}). Basically, the saturation scale in the nucleus grows with the quotient of the transverse parton densities to the power $\epsilon$, where $\epsilon = 1.27$ as determined by the fit to the lepton-nucleus data \cite{Armesto_scal}. As an example, we get for $W\gg M_Z$ the coherent $Z$ production cross sections $\sigma_{tot}(\gamma Pb\rightarrow Z^0 Pb)\approx 0.16\,  fb \,(W/\mathrm{GeV})^{1.8}$ and $\sigma_{tot}(\gamma Ca\rightarrow Z^0 Ca)\approx 1.2\cdot 10^{-2}\, fb\,(W/\mathrm{GeV})^{1.9}$, respectively. For the nuclear dependence, due to the geometric scaling property it can be shown that $\sigma (\gamma A\rightarrow Z^0 A)\propto [\pi R_A^2Q_{\mathrm{sat},A}^2]^2/B \approx A^{\frac{2}{3}(1+\epsilon)} \,W^{\,4\lambda}$, where $B\propto R_A^2$ is the $t$-slope for the nuclear scattering. This is consistent with no absorption, as referred before.
In what follows we will use the calculation presented above to compute the diffractive $Z^0$ photoproduction in $pp$ and $AA$ collisions. In these reactions, the rapidity distribution for $Z$ production is directly related to cross section discussed here.

\begin{table}[t]
\begin{ruledtabular}
\begin{tabular}{||c|c|c||}
$h_1 h_2$  & Total    &  $|y| < 1$ \\
\hline
$\mathrm{PbPb}$ (${\cal L}=0.42$ mb$^{-1}$s$^{-1}$) &  0.3 nb ($0.13 \times 10^{-6}$ s$^{-1}$) &  0.25 nb \\
\hline
$\mathrm{CaCa}$ (${\cal L}=43$ mb$^{-1}$s$^{-1}$) &  10 pb ($0.45 \times 10^{-6}$ s$^{-1}$) & 8 pb \\
\hline
$pp$ (${\cal L}=10^{7}$ mb$^{-1}$s$^{-1}$)  &  11 fb ($1.1 \times 10^{-4}$ s$^{-1}$)  & 1.9 fb\\
\end{tabular}
\end{ruledtabular}
\caption{\label{tab:table1}  The integrated cross section (events rates/second) for the diffractive $Z^0$ photoproduction in $pp/CaCa/PbPb$ collisions at LHC energies. }
\end{table}

\section{Photoproduction of $Z^0$ in coherent collisions}

Lets consider the hadron-hadron interaction at large impact parameter ($b > R_{h_1} + R_{h_2}$) and at ultra relativistic energies. In this regime we expect the electromagnetic interaction to be dominant.
In  heavy ion colliders, the heavy nuclei give rise to strong electromagnetic fields due to the coherent action of all protons in the nucleus, which can interact with each other. In a similar way, it also occurs when considering ultra relativistic  protons in $pp(\bar{p})$ colliders \cite{upcs}.
The cross section for the diffractive $Z^0$ photoproduction in a coherent  hadron-hadron collision is  given in Eq. (\ref{sigAA}). For related works in coherent interactions see Ref. \cite{vicmag}.  Considering the requirement that  photoproduction
is not accompanied by hadronic interaction (ultraperipheral
collision) an analytic approximation for the equivalent photon flux of a nuclei can be calculated, which is given by \cite{upcs}
\begin{eqnarray}
\frac{dN_{\gamma}\,(\omega)}{d\omega}= \frac{2\,Z^2\alpha_{em}}{\pi\,\omega}\, \left[\bar{\eta}\,K_0\,(\bar{\eta})\, K_1\,(\bar{\eta})+ \frac{\bar{\eta}^2}{2}\,{\cal{U}}(\bar{\eta}) \right]\,
\label{fluxint}
\end{eqnarray}
where
 $\omega$ is the photon energy,  $\gamma_L$ is the Lorentz boost  of a single beam and $\eta
= \omega b/\gamma_L$; $K_0(\eta)$ and  $K_1(\eta)$ are the
modified Bessel functions.
Moreover, $\bar{\eta}=2\omega\,R_A/\gamma_L$ and  ${\cal{U}}(\bar{\eta}) = K_1^2\,(\bar{\eta})-  K_0^2\,(\bar{\eta})$.
The Eq. (\ref{fluxint}) will be used in our calculations of $Z$ production in $AA$ collisions. On the other hand, for   proton-proton interactions, we assume that the  photon spectrum is given by Ref. \cite{Dress}.
The coherence condition limits the photon virtuality to very low values, which implies that for most purposes, they can be considered as real. Moreover, if we consider $pp/PbPb$ collisions at LHC, the Lorentz factor  is
$\gamma_L = 7455/2930 $, giving the maximum c.m.s. $\gamma N$ energy
$W_{\gamma p} \approx 8390/950$ GeV.   Finally, since photon emission is coherent over the entire nucleus and the photon is colorless we expect that the events to be characterized by   two   rapidity gaps. Therefore, we have that distinctly from central collisions, in ultraperipheral collisions the diffractive $Z^0$ production should occur in a clean environment where the hadronic background is reduced.

\begin{figure*}[t]
\begin{tabular}{cc}
\includegraphics[scale=0.78]{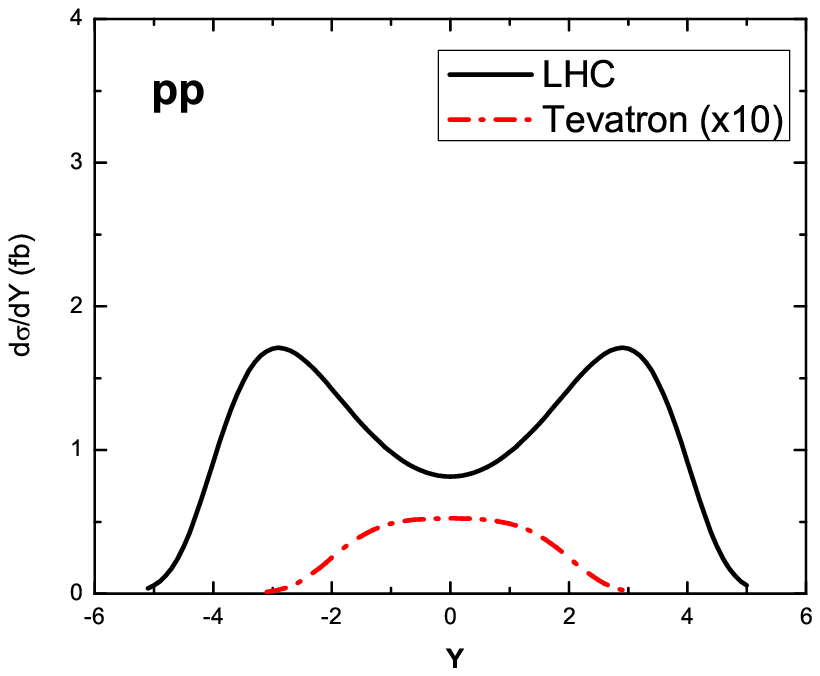} &
\includegraphics[scale=0.78]{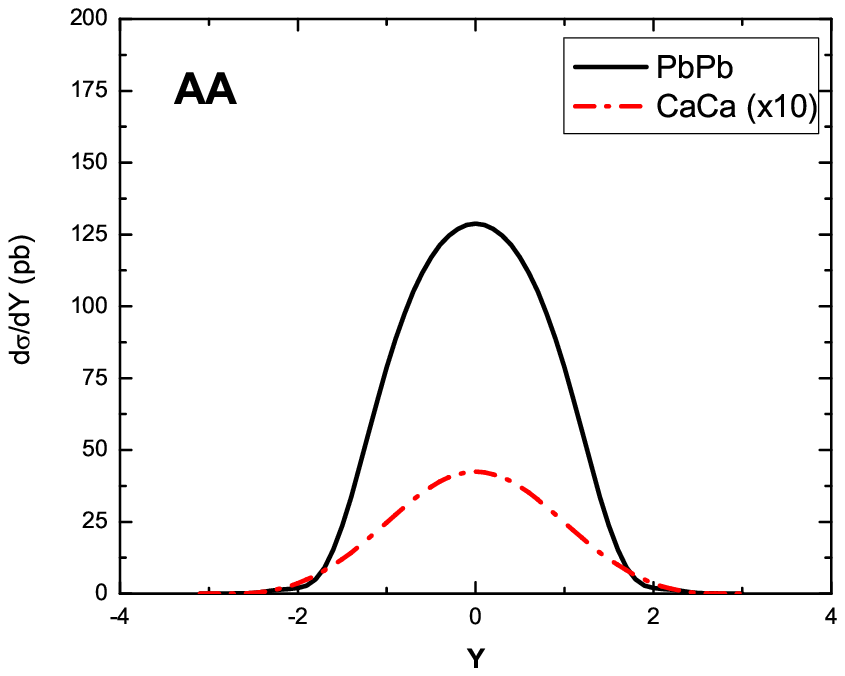}
\end{tabular}
 \caption{(Color online) The rapidity distribution for diffractive $Z^0$ photoproduction in $pp$  and $AA$  collisions (See text).}
\label{fig:2}
\end{figure*}

Lets calculate the rapidity distribution and total cross sections for $Z^0$ boson production in  coherent hadron-hadron collisions, which is given by
\begin{eqnarray}
\frac{d\sigma \,\left[h_1 + h_2 \rightarrow   h_1 \otimes Z^0 \otimes h_2 \right]}{dY} = \omega \frac{dN_{\gamma} (\omega )}{d\omega }\,\sigma_{\gamma h \rightarrow Z^0 h}\left(\omega \right)\,
\label{dsigdy}
\end{eqnarray}
where $Y\propto \ln \, (2 \omega/m_{Z^0})$ and  $\otimes$ represents the presence of a rapidity gap.
Consequently, given the photon flux, the rapidity distribution is thus a direct measure of the photoproduction cross section for a given energy. In Fig. \ref{fig:2} we present respectively our predictions for $Z^0$ production in coherent $pp$ (left panel) and $AA$ (right panel) collisions. For the $pp$ case, energies of Tevatron and LHC are considered. For $AA$ collisions, we consider only LHC energy and compute the cross section for light (Ca) and heavy (Pb) nuclei. We summarize the integrated cross sections in Table 1 for the LHC case, considering integration over all rapidity and the conservative range $|y|\leq 1$. For completeness, we quote the results for the $p\bar{p}$ Tevatron energy: for total rapidity one has 0.21 fb and 0.12 fb for rapidity $|y|\leq 1$. Assuming ${\cal L}= 2 \times 10^{5}$ mb$^{-1}$s$^{-1}$ at Tevatron, it implies that the corresponding events rates/second is $0.4 \times 10^{-7}$ s$^{-1}$.
These  results indicate that the experimental analyzes of this process could be feasible only at the LHC for the $pp$ mode. However, a more detailed discussion about the identification of the outgoing $Z^0$ and its main backgrounds is necessary. One has that the main $Z^0$ decay mode is the hadronic one, but it is expected to be difficult to identify, due to the background from photoproduction of dijets. The magnitude of the cross section of this process in coherent interactions still is an open question, as well as its transverse momentum distribution. It needs to be studied before one can assess the observability of hadronic $Z^0$ decays. In other words, a comparison between the characteristics of the two final states deserves more detailed studies, which we postpone for a future publication.
Consequently, in principle, the experimental search of the $Z^0$ would have to rely on the clean leptonic $Z^0$ decay modes ($Z^0 \rightarrow l^+ l^-$). It implies the suppression of the rate by a factor $0.7$ (6.7 \%  leptonic branching ratio, i.e. $Z\rightarrow e^+e^-,\,\mu^+\mu^-$), which reduces in approximately one order of magnitude  the values presented in the Table \ref{tab:table1}. This gives $\approx 100$ events/year in $pp$ at the LHC after including branching.  Consequently, if we only consider the leptonic decays of the $Z^0$, the separation of this process in coherent $pp$ interactions is almost impossible. Finally, the scenario for the $AA$ mode is pessimistic, giving very small event rates. This case also presents the additional disadvantage of having smaller running time than the $pp$ case, which diminishes the event rates per year.

Lets now discuss some of the main backgrounds. The $Z^0$ bosons can also be produced in inclusive hadron - hadron interactions. In comparison with inclusive $Z^0$ production, which is characterized by the process $h_1 + h_2 \rightarrow Z^0 +X$, one have that the photoproduction cross section is smaller by approximately three order of magnitudes for proton-proton collisions. For nuclear collisions, this factor is smaller due to presence of shadowing corrections (See e.g. Ref. \cite{vogt}). Although the photoproduction cross section would be a small factor of the hadronic cross section, the separation of this channel is feasible if we impose the presence of two rapidity gaps in the final state. This should eliminate almost all of the hadroproduction events while retaining most of the photoproduction interactions. However, two rapidity gaps in the final state can also be generated in  hadron - hadron interactions via the $WW$ fusion, but in this case it are accompanied by two forward jets which can be used to discriminate between the two processes. Moreover, the processes mediated by a photon implies a distinct transverse momentum distribution, which can be used to separate the diffractive photoproduction \cite{upcs}. Another background process is the production of electron - positron pairs. Due to their small mass, they are produced copiously, predominantly at low invariant mass and energies and in the forward and backward direction \cite{baur2}. In contrast, in the leptonic $Z^0$ decays, the pair have an invariant mass equal to $Z^0$ mass and hence very large individual transverse momenta.



\section{Summary}

The study of the $Z^0$ production and related physics has contributed  significantly for the understanding of the Standard Model as well as for the search of new physics. At LHC energies it should be produced copiously in $pp$ and $AA$ collisions. However, due to the overwhelming   background of QCD multi-jet production, it is almost to impossible to study the physics of hadronic $Z^0$ decays in central collisions.  In this letter we propose to study the diffractive $Z^0$ photoproduction in coherent hadron - hadron interactions, which is characterized by clean environment where  the $Z^0$ boson produced is separated of the outgoing hadrons by rapidity gaps. This process could be tested using  the forward proton detectors proposed to be installed in the CMS and ATLAS collaborations at LHC. We have generalized the description of the DVCS process for $Z^0$ production and estimated the  total cross section for the exclusive process $\gamma^* h\rightarrow Z^0 h$ ($h=p,A$) for different  energies, photon virtualities and atomic numbers.  Finally, we have calculated the
$h\,h\rightarrow h\, Z^0 \,h$ process and obtained that the experimental analyzes of this process using the leptonic $Z^0$ decay mode is a hard task in coherent hadron-hadron collisions at LHC. The present results are consistent with similar studies \cite{motyka} using the color dipole approach. The feasibility seems to be worst in ultraperipheral heavy ion collisions. However, if were possible to separate the hadronic $Z^0$ decay from the background of dijet photoproduction, the analyzes of exclusive $Z^0$ production  could be feasible in $pp$ collisions at the LHC.

\begin{acknowledgments}
 The authors are grateful to Mike Albrow, David d\'\,Enterria, L. Motyka and G. Watt for useful comments.
This work was financed by the Brazilian funding
agencies CNPq and FAPERGS.

\end{acknowledgments}


\begin{thebibliography}{99}

\bibitem{ellis}
 J.~Ellis,
Acta Phys.\ Polon.\  B {\bf 38}, 1071 (2007);  Nature {\bf 448}, 297 (2007).


\bibitem{forshaw}
  J.~R.~Forshaw,
  PoS  {\bf DIFF2006}, 055 (2006)
  [arXiv:hep-ph/0611274].

\bibitem{reviewZ0}
  M.~W.~Grunewald,
  arXiv:0710.2838 [hep-ex].

\bibitem{martin}
  A.~D.~Martin, R.~G.~Roberts, W.~J.~Stirling and R.~S.~Thorne,
  Eur.\ Phys.\ J.\  C {\bf 14}, 133 (2000)

  \bibitem{upcs}
 G. Baur, K. Hencken, D. Trautmann, S. Sadovsky, Y. Kharlov, Phys.
Rep. {\bf 364}, 359 (2002);
 C.~A. Bertulani, S.~R.~Klein and J.~Nystrand, Ann. Rev. Nucl. Part. Sci. {\bf 55}, 271 (2005).

\bibitem{detec}
A. De Roeck, in J. Bartels et al., arXiv:0712.3633 [hep-ph], p. 181; M. Tasevsky, ibid., p.145; A. Hamilton, ibid, p.160.




\bibitem{dipole}  N.~N.~Nikolaev and B.~G.~Zakharov, Z. Phys. C{\bf 49}, 607
(1991); Z. Phys. C {\bf 53}, 331 (1992); A.~H.~Mueller, Nucl. Phys. B
{\bf 415}, 373 (1994)


\bibitem{dvcs}
M.~McDermott, R.~Sandapen and G.~Shaw,
  Eur.\ Phys.\ J.\  C {\bf 22}, 655 (2002);  L.~Favart and M.~V.~T.~Machado,
  Eur.\ Phys.\ J.\  C {\bf 29}, 365 (2003); Eur.\ Phys.\ J.\  C {\bf 34}, 429 (2004).


\bibitem{fss}
J.~R.~Forshaw, R.~Sandapen and G.~Shaw,
JHEP {\bf 0611}, 025 (2006).

\bibitem{FM}
L.~Favart and M.~V.~T.~Machado,
Eur.\ Phys.\ J.\ C {\bf 29}, 365 (2003); Eur.\ Phys.\ J.\ C {\bf 34}, 429 (2004).


\bibitem{MPS}
C.~Marquet, R.~Peschanski and G.~Soyez,
  Phys.\ Rev.\  D {\bf 76}, 034011 (2007)


\bibitem{kowtea}
H. Kowalski and D. Teaney,
{ Phys. Rev. D} {\bf 68}, 114005 (2003).

\bibitem{bartels_non}
  J.~Bartels, K.~J.~Golec-Biernat and K.~Peters,
  Acta Phys.\ Polon.\  B {\bf 34}, 3051 (2003)



\bibitem{hdqcd}
E.~Iancu and R.~Venugopalan,
arXiv:hep-ph/0303204;
  H.~Weigert,
  Prog.\ Part.\ Nucl.\ Phys.\  {\bf 55}, 461 (2005); J.~Jalilian-Marian and Y.~V.~Kovchegov, Prog.\ Part.\ Nucl.\ Phys.\  {\bf 56}, 104 (2006).





\bibitem{Iancu:2003ge}
  E.~Iancu, K.~Itakura and S.~Munier,
  Phys.\ Lett.\ B {\bf 590}, 199 (2004).



\bibitem{vacca}
  K.~Peters and G.~P.~Vacca,
  Eur.\ Phys.\ J.\  C {\bf 30}, 345 (2003)



\bibitem{motyka} L. Motyka and G. Watt, Phys. Rev. {\bf D78}, 014023 (2008).
\bibitem{motyka_disc} We thank L. Motyka for useful discussions concerning this issue.
\bibitem{BartelsLoewe} J. Bartels and M. Loewe, Z. Phys. {\bf C12}, 263 (1982).

\bibitem{Pumplin} J. Pumplin, arXiv:9612356 [hep-ph].

\bibitem{Diehl} E.R. Berger, M. Diehl and B. Pire, Eur.\ Phys.\ J.\  C {\bf 23}, 675 (2002).

\bibitem{Armesto_scal}
  N.~Armesto, C.~A.~Salgado and U.~A.~Wiedemann,
  Phys.\ Rev.\ Lett.\  {\bf 94}, 022002 (2005).




\bibitem{vicmag}
  V.~P.~Goncalves and M.~V.~T.~Machado,
  Phys.\ Rev.\  D {\bf 77}, 014037 (2008); Phys.\ Rev.\  D {\bf 75}, 031502 (2007); Phys.\ Rev.\  C {\bf 73}, 044902 (2006); Eur.\ Phys.\ J.\  C {\bf 40}, 519 (2005); Phys.\ Rev.\  D {\bf 71}, 014025 (2005); Eur.\ Phys.\ J.\  C {\bf 31}, 371 (2003).

\bibitem{Dress}
M.~Drees and D.~Zeppenfeld, Phys.\ Rev.\ D {\bf
39}, 2536 (1989).

\bibitem{vogt}
  R.~Vogt,
  Phys.\ Rev.\  C {\bf 64}, 044901 (2001).

\bibitem{baur2}
  G.~Baur, K.~Hencken and D.~Trautmann,
  Phys.\ Rept.\  {\bf 453}, 1 (2007).

\end{thebibliography}
\end{document}